\documentclass{article}
\usepackage{epsfig}

\makeatletter
\@addtoreset{equation}{section}

\makeatother

\begin{document}

\begin{flushright}
Sep 2009

SNUTP09-014
\end{flushright}

\begin{center}

\vspace{5cm}

{\LARGE 
\begin{center}
Notes on Matter in Horava-Lifshitz Gravity
\end{center}
}

\vspace{2cm}

Takao Suyama \footnote{e-mail address : suyama@phya.snu.ac.kr}

\vspace{1cm}

{\it Center for Theoretical Physics, 

Seoul National University, 

Seoul 151-747 Korea}

\vspace{3cm}

{\bf Abstract} 

\end{center}

We investigate the dynamics of a scalar field governed by the Lifshitz-type action which should appear naturally in 
Horava-Lifshitz gravity. 
The wave of the scalar field may propagate with any speed without an upper bound. 
To preserve the causality, the action cannot have a generic form. 
Due to the superluminal propagation, a formation of a singularity may cause the breakdown of the predictability of the theory. 
To check whether such a catastrophe could occur in Horava-Lifshitz gravity, we investigate the dynamics of a dust. 
It turns out that the dust does not collapse completely to form a singularity in a generic situation, 
but expands again after it attains a maximum energy density.

\newpage

\vspace{1cm}

\section{Introduction}

\vspace{5mm}

The construction of a quantum theory of gravity is one of the long-standing problems in theoretical physics. 
The main difficulty is the non-renormalizability which can be easily seen from the fact that the Newton constant has a negative 
mass dimension. 
Recent investigations (see \cite{Bern} and references therein) 
on the perturbative amplitudes in supergravity theories seem to indicate that there would exist an 
unknown 
mechanism of cancellation among diagrams which might make some supergravity theories renormalizable. 
There is another proposal, called Horava-Lifshitz gravity \cite{Horava:2008ih}\cite{Horava:2009uw}, 
which has attracted much attention recently. 
The basic idea of the approach is to modify the UV behavior of general relativity so that the perturbative renormalizability 
can be realized. 
The price to pay for this achievement is the abandonment of the Lorentz symmetry in the high energy regime: in this context, 
the Lorentz symmetry is regarded as an 
approximate symmetry observed only at low energy. 
See also 
\cite{Orlando:2009en}\cite{Orlando:2009az}\cite{Kluson:2009sm}\cite{Kluson:2009hr}\cite{Kluson:2009rk}\cite{Calcagni:2009qw}.

Since the Lorentz symmetry does not exist in the high energy regime, the notion of causality should be largely modified. 
For example, since the dispersion relation of the propagating modes is modified, they can propagate with any speed. 
This seems to raise a puzzle if this superluminal propagation is realized in the presence of a singularity. 
It would be natural to expect that a singularity would produce particles with an arbitrarily high energy. 
If they are produced, they can propagate with arbitrarily high speed, and therefore, there would always exist 
a particle which 
can reach an observer at infinity. 
Since such a particle would carry information of the singularity, the possibility of this phenomenon shows that the 
predictability of the theory breaks down. 
In general relativity, it is believed that any singularity should be hidden behind the event horizon, and it cannot affect any 
physical phenomena outside the event horizon. 
However, it does not seem to be the case in Horava-Lifshitz gravity since the light cone structure is not relevant, and 
therefore the ``event horizon'' defined as in general relativity cannot hide the singularity. 
The black hole solutions in Horava-Lifshitz gravity were investigated in 
\cite{Lu:2009em}\cite{Nastase:2009nk}\cite{Cai:2009pe}\cite{Cai:2009ar}\cite{Colgain:2009fe}\cite{Myung:2009dc}\cite{Kehagias:2009is}\cite{Cai:2009qs}\cite{Ghodsi:2009rv}\cite{Myung:2009va}\cite{Chen:2009gs}\cite{Park:2009zra}\cite{Ghodsi:2009zi}\cite{Castillo:2009ci}\cite{Peng:2009uh}\cite{Lee:2009rm}\cite{Myung:2009us}\cite{Konoplya:2009ig}\cite{Kim:2009dq}. 
The other subtle issues in Horava-Lifshitz gravity were pointed out in 
\cite{Li:2009bg}\cite{Charmousis:2009tc}\cite{Blas:2009yd}\cite{Bogdanos:2009uj}. 
The dynamics of an anisotropic classical mechanical system was discussed recently \cite{Romero:2009qs}. 

In this paper, we investigate the propagation of a scalar field whose action is of the Lifshitz-type. 
We employ the WKB approximation, and study the trajectory of the ray of the wave of the scalar field. 
As is expected, it turns out that the wave can propagate with an arbitrary speed. 
Because of this, the light cone structure cannot have a physical relevance. 
Therefore, in Horava-Lifshitz gravity coupled to matter, there might occur the breakdown of the predictability, as pointed out 
above. 
To check this, we investigate a gravitational collapse in Horava-Lifshitz gravity, and check whether a singularity would be 
formed. 
We choose the simplest situation: the collapse of a dust preserving the homogeneity and the isotropy. 
Naively, this would be the most probable situation for a singularity to form\footnote{
In this analysis, we assume, just for simplicity, that the equation of state is always the same. 
It may be expected \cite{Rama:2009xc} 
that the equation of state may change with the energy scale, but the inclusion of this property to the 
analysis will not be compatible with the homogeneity and the isotropy, and therefore, the analysis will become complicated. 
It will be very interesting to study this issue further. 
}. 
Interestingly enough, it turns out that the dust does not collapse completely in a generic situation. 
Instead, at a maximum energy density, the dust ceases to collapse, and then starts expanding. 
The repulsive force necessary for this behavior is induced by the higher derivative terms in Horava-Lifshitz gravity which are 
necessary to modify the UV behavior. 
Note that the analysis may be closely related to 
\cite{Brandenberger:2009yt}\cite{Carloni:2009jc}\cite{Wang:2009az}\cite{Wang:2009yz}\cite{Calcagni:2009ar}\cite{Saridakis:2009bv}\cite{Cai:2009in}\cite{Leon:2009rc} 
some of which discussed a bounce cosmology in Horava-Lifshitz gravity. 
It should be noted that the analysis in this paper cannot be a proof of the absence of singularities in Horava-Lifshitz theory. 
The investigation on situations more general than the dust collapse is beyond the scope of this paper. 

This paper is organized as follows. 
In section 1, we investigate the propagation of a scalar field governed by a Lifshitz-type action, using the WKB approximation. 
Section 2 argues the issues related to causality. 
A gravitational collapse in Horava-Lifshitz gravity is discussed in section 3. 
Section 4 is devoted to discussion.

\vspace{1cm}

\section{Geometric optics for Horava-Lifshitz scalar}

\vspace{5mm}

We consider a complex scalar field whose dynamics is governed by the action 
\begin{equation}
S = \int d^3xdt\sqrt{g}N\left[ \frac1{N^2}|D_t\phi|^2-f(\phi,\phi^\dag,\nabla,g) \right], 
   \label{HLscalar}
\end{equation}
where 
\begin{equation}
D_t\phi := \partial_t\phi-N^i\partial_i\phi. 
\end{equation}
and $i$ runs from 1 to 3. 
The fields $N,N^i,g_{ij}$ are combined to form a four-dimensional metric 
\begin{equation}
ds^2 = -N^2dt^2+g_{ij}(dx^i+N^idt)(dx^j+N^jdt). 
   \label{4dim}
\end{equation}
The expression $f(\phi,\phi^\dag,\nabla,g)$ is a scalar with respect to the three-dimensional diffeomorphisms. 
More explicitly, $f(\phi,\phi^\dag,\nabla,g)$ is assumed to have the form 
\begin{equation}
f(\phi,\phi^\dag,\nabla,g) = \sum_{n=0}^MG^{i_1\cdots i_nj_1\cdots j_n}\nabla_{i_1}\cdots\nabla_{i_n}\phi^\dag\nabla_{j_1}\cdots
 \nabla_{j_n}\phi,
\end{equation}
where $G^{i_1\cdots i_nj_1\cdots j_{n}}$ is a three-dimensional tensor constructed from $g^{ij}$, and $\nabla_i$ is the covariant 
derivative with respect to $g_{ij}$. 
In the following, $N$ is assumed to be independent of the spatial coordinates $x^i$. 

The equation of motion of $\phi$ is 
\begin{equation}
-\frac1{\sqrt{g}N}\partial_t\left[ \sqrt{g}N\frac1{N^2}D_t\phi \right]+\frac1{N^2}\nabla_i(N^iD_t\phi)
 -\sum_{n=0}^M (-1)^nG^{i_1\cdots i_nj_1\cdots j_{n}}\nabla_{i_n}\cdots\nabla_{i_1}\nabla_{j_1}\cdots\nabla_{j_{n}}\phi = 0. 
\end{equation}
It may be possible to analyze the propagation of $\phi$ directly. 
However, the analysis will be easier if we restrict ourselves to deal with ``geometric optics'', that is, 
a particle mechanical description of the ray of $\phi$. 
We use WKB approximation to obtain such a classical mechanical system. 
Assuming $\phi=e^{iS}$ with $S$ real, and neglecting sub-leading terms, we obtain 
\begin{equation}
\frac1{N^2}(D_tS)^2-\sum_{n=0}^M G^{i_1\cdots i_nj_1\cdots j_n}(\partial_{i_1}S)\cdots(\partial_{i_n}S)(\partial_{j_1}S)\cdots 
 (\partial_{j_n}S) = 0. 
\end{equation}
The left-hand side can be regarded as a Hamiltonian $H$ by replacing $\partial_\mu S$ with $p_\mu$. 
Then the equations of motion can be obtained as the Hamilton's equations. 
For example, if we started with the relativistic action, then the resulting Hamiltonian would be 
\begin{equation}
H_{\rm rel} = -g^{\mu\nu}p_\mu p_\nu-m^2, 
   \label{rel}
\end{equation}
and the Hamilton's equations with this Hamiltonian 
\begin{equation}
\frac{dx^\mu}{d\tau} = -2g^{\mu\nu}p_\nu, \hspace{1cm} \frac{dp_\mu}{d\tau} = \partial_\mu g^{\rho\sigma}p_\rho p_\sigma, 
\end{equation}
are equivalent to the geodesic equation with respect to the metric $g_{\mu\nu}$. 
Note that the geodesic equation can also be derived from the conservation law of the energy-momentum tensor. 

For the general case, the Hamiltonian is 
\begin{equation}
H = \frac1{N^2}(p_t-N^ip_i)^2-h(p^2), 
   \label{nonrel}
\end{equation}
where $p^2=g^{ij}p_ip_j$ and $h(x)$ is a polynomial defined as 
\begin{equation}
h(p^2) = f(1,1,p,g). 
\end{equation}
For the small momentum region, we expect that the system is approximately relativistic. 
This implies 
\begin{equation}
h(p^2) = m^2+p^2+O(p^4). 
\end{equation}
We require that the (Euclidean) path-integral for this system should be well-defined. 
This implies 
\begin{equation}
\lim_{p^2\to\infty}h(p^2) = +\infty. 
\end{equation}

The Hamilton's equations are 
\begin{eqnarray}
\frac{dx^i}{d\tau} &=& -\frac2{N^2}(p_t-N^jp_j)N^i-2h'(p^2)g^{ij}p_j, \\
\frac{dt}{d\tau} &=& \frac2{N^2}(p_t-N^ip_i), \\
\frac{dp_i}{d\tau} &=& \frac2{N^2}(p_t-N^jp_j)p_k\partial_iN^k+h'(p^2)\partial_ig^{kl}p_kp_l, \\
\frac{dp_t}{d\tau} &=& \frac2{N^2}(p_t-N^jp_j)p_k\partial_tN^k+\frac2{N^3}\partial_tN(p_t-N^ip_i)^2
                       +h'(p^2)\partial_tg^{kl}p_kp_l. 
\end{eqnarray}
This shows that a ray of $\phi$ does not follow any geodesic defined by the metric (\ref{4dim}). 

In general, it is very difficult to eliminate the momenta using the velocities in the above equations. 
Let us consider a simple case in which the $h(x)$ is a monomial, that is, 
\begin{equation}
H_n = \frac1{N^2}(p_t-N^ip_i)^2-\frac{c}{2n}(p^2)^n
\end{equation}
with a positive $c$. 
In this case, we obtain 
\begin{eqnarray}
v^t &=& \frac2{N^2}(p_t-N^ip_i), \\
v^i+N^iv^t &=& -c(p^2)^{n-1}g^{ij}p_j. 
\end{eqnarray}
The momenta are given explicitly in terms of the velocities, and we obtain the Lagrangian 
\begin{eqnarray}
L &=& v^tp_t+v^ip_i-H \nonumber \\
  &=& \frac{N^2}4(v^t)^2-c'(V^2)^{\frac n{2n-1}}, 
   \label{Lagrangian}
\end{eqnarray}
where 
\begin{equation}
c' = \left( 1-\frac1{2n} \right)c^{-\frac1{2n-1}}, 
\end{equation}
and 
\begin{equation}
V^2 = g_{ij}V^iV^j, \hspace{5mm} V^i = v^i+N^iv^t. 
\end{equation}

\vspace{1cm}

\section{Causality}

\vspace{5mm}

Since the Hamiltonian (\ref{nonrel}) is different from the relativistic one (\ref{rel}), 
the equations of motion are different from the geodesic equation for the metric (\ref{4dim}). 
Therefore, the issue on the causality would be very different from the familiar relativistic one. 

In our setup, the existence of the absolute time is assumed. 
This implies that the presence of closed time-like trajectories is forbidden. 
In this case, a typical causality-violating trajectory is the one which at first moves forward in time but at a later time moves 
backward in time. 
In other words, there could be a closed trajectory. 
Such a trajectory is possible only if $\frac{dt}{d\tau}=0$ is realizable. 
This implies that $p_t-N^ip_i=0$, and therefore $h(p^2)=0$ must be realized at some value of $\tau$. 
As a result, such a causality violation never occurs if $h(p^2)$ is always positive. 
(This is of course the case for the relativistic system.) 
On the other hand, if $h(p^2)$ allows to have a zero, then, in general, there would exist a trajectory which ``turns around'' at 
some value of $t$ since $\frac{d^2t}{d\tau^2}$ is non-zero in general at the time when the trajectory turns around. 

Let us examine the simple case in which $N=1$ and $N^i=0$. 
In this case, a trajectory turns around when $p_t=0$, and at this moment 
\begin{equation}
\frac{d^2t}{d\tau^2} = 2\frac{dp_t}{d\tau} = 2h'(p^2)\partial_tg^{kl}p_kp_l. 
\end{equation}
The value of the right-hand side depends on the background metric, and is generically non-zero unless the background is static. 

There is another possibility of a strange behavior which never happens in the relativistic system. 
One notices that the expressions in the right-hand side of the Hamilton's equations are simplified when $h'(p^2)=0$ is realized. 
At this point in the phase space, the correspondence between the momenta and the velocities is lost. 
For example, consider again the case $N=1$ and $N^i=0$. 
Then, the $\tau$-derivatives of $x^i$, $p_i$ and $p_t$ all vanish when $h'(p^2)=0$, and $\frac{dt}{d\tau}=2p_t$. 
In addition, one can show that 
\begin{equation}
\frac{d}{d\tau}p^2 = 2p_t\partial_tg^{kl}p_kp_l. 
\end{equation}
Therefore, if the background is static, then $p^2$ and also $p_t$ are constant, due to the conservation of the Hamiltonian, and 
\begin{eqnarray}
p_\mu = \mbox{const.} \hspace{5mm} x^i = \mbox{const.} \hspace{5mm} t = 2p_t\tau+t_0
\end{eqnarray}
is a solution. 
It looks very strange since non-zero $p_i$ correspond to the trajectory at rest as long as $h'(p^2)=0$. 
Note that in a generic background, since $\frac{d}{d\tau}p^2$ would not be zero when $h'(p^2)=0$, and therefore, such a strange 
solution is not allowed. 

In the following, let us assume that 
$h(x)$ is a monotonically increasing function with $h'(x)>0$ 
in order to avoid the strange behaviors of the trajectory mentioned above. 
In this case, the causality is kept in the sense that there is no trajectory which moves backward in time, 
and therefore, no closed trajectory. 
However, the notion of the causality is 
still 
quite different from the one in general relativity. 
In particular, 
as is easily anticipated from the action (\ref{HLscalar}) with which we started, there is no 
physical relevance of the light cone. 
In fact, one can calculate the relativistic norm of the velocity as follows: 
\begin{eqnarray}
g_{\mu\nu}v^\mu v^\nu 
&=& -N^2(v^t)^2+g_{ij}(v^i+N^iv^t)(v^j+N^jv^t) \nonumber \\
&=& 4[ -h(p^2)+p^2(h'(p^2))^2 ]. 
\end{eqnarray}
This norm is constant only when the action (\ref{HLscalar}) is relativistic. 
This becomes indefinitely large and positive when the spatial momenta $p_i$ become large. 
This implies that a trajectory which was time-like at first may becomes space-like, and therefore, the trajectory 
can escape from the light cone defined by the metric (\ref{4dim}) in the usual manner. 
Note that this property of the trajectories is inevitable as long as the degree of $h(x)$ is larger than 1, that is, as long as 
the UV behavior is modified. 
In contrast, the presence of a closed trajectory, for example, can be avoided by choosing a suitable functional form of $h(x)$. 

The irrelevance of the light cone poses the following question: is the formation of a black hole in Horava-Lifshitz gravity 
allowed? 
Suppose that black holes are defined in the same manner as in general relativity. 
In Horava-Lifshitz gravity, the event horizon cannot hide the singularity inside, since there exist trajectories which can 
escape from the inside of the event horizon. 
Therefore, it seems that the predictability of the theory would break down whenever a singularity is formed. 
In the next section, we will investigate the collapse of a dust in Horava-Lifshitz gravity. 
We choose this situation since, due to the absence of the pressure, the formation of a singularity by the collapse of the dust 
seems to be easily realized. 
Interestingly enough, it will be shown that the higher derivative terms prevent the dust from collapsing completely, and the 
dust will finally form a compact object with a finite size.

\vspace{1cm}

\section{Dynamics of dust in Horava-Lifshitz gravity}

\vspace{5mm}

Let us first recall the analysis of the dust collapse in general relativity \cite{Weinberg}. 
Since Horava-Lifshitz gravity typically includes a cosmological constant, we include it to the analysis in \cite{Weinberg} 
so that the effects of the higher derivative terms may be highlighted. 

We consider the following action 
\begin{equation}
S = \int d^4x\sqrt{-g}\left[ \frac2{\kappa^2}R+\sigma+L_m \right], 
\end{equation}
where $\kappa^2=32\pi G$. 
For simplicity, we consider the collapse which preserves the homogeneity and the isotropy. 
Then the metric is of the Robertson-Walker form 
\begin{eqnarray}
ds^2 &=& -dt^2+a(t)^2h_{ij}(x)dx^idx^j, \nonumber \\
h_{ij}(x)dx^idx^j &=& \frac{dr^2}{1-kr^2}+r^2(d\theta^2+\sin^2\theta d\phi^2). 
\end{eqnarray}
Let $\rho(t)$ be the energy density of the dust. 
The conservation law determines $\rho=\rho_0a^{-3}$. 
The Einstein equation reduces to 
\begin{equation}
\left( \frac{da}{dt} \right)^2 = \frac{\kappa^2}{12}\left[ \frac{\rho_0}a-\sigma a^2 \right]-k, 
   \label{relF}
\end{equation}
where we rescaled the radial coordinate $r$ so that $a(0)=1$ is satisfied. 
We assume that the collapse starts at $t=0$, that is, we choose the initial condition $\dot{a}(0)=0$, which fixes 
\begin{equation}
k = \frac{\kappa^2}2(\rho_0-\sigma). 
\end{equation}
In the case $\sigma=0$, there is an exact solution of (\ref{relF}). 
The solution is a cycloid whose parametric expression is 
\begin{eqnarray}
a(\psi) &=& \frac12(1+\cos\psi), \\
t(\psi) &=& \frac1{2\sqrt{k}}(\psi+\sin\psi). 
\end{eqnarray}
The collapse $a=0$, at which $\rho$ diverges, occurs at a finite time $t=T$ where 
\begin{equation}
T = \frac{\pi}\kappa\sqrt{\frac3{\rho(0)}}. 
\end{equation}
The behavior of $a$ near $a=0$ is similar to the above exact solution even for a nonzero $\sigma$ since its contribution is 
negligible when $a$ is small. 

\vspace{5mm}

Next, we consider the same situation in the context of Horava-Lifshitz gravity. 
Following \cite{Kiritsis:2009sh}, we consider a rather general action 
\begin{eqnarray}
S &=& \int d^3xdt\sqrt{g}N\Bigl[ \alpha(K_{ij}K^{ij}-\lambda K^2)+\beta C_{ij}C^{ij}+\gamma{\cal E}^{ijk}R_{il}\nabla_jR^l{}_l
        \nonumber \\
  & & \hspace*{2.5cm}+\zeta R_{ij}R^{ij}+\eta R^2+\xi R+\sigma +L_m \Bigr], 
         \label{HLgravity}
\end{eqnarray}
where $L_m$ is the matter Lagrangian. 
Various quantities are defined as  
\begin{eqnarray}
K_{ij} &=& \frac1{2N}(\dot{g}_{ij}-\nabla_iN_j-\nabla_jN_i), \\
K &=& K_{ij}g^{ij}, \\
C^{ij} &=& \frac{\epsilon^{ikl}}{\sqrt{g}}\nabla_k\left( R^j{}_l-\frac14R\delta^j{}_l \right), 
\end{eqnarray}
and $R_{ij}$, $R$ are the Ricci tensor and the scalar curvature with respect to $g_{ij}$. 

As in the relativistic case, 
let us consider a dynamics of the dust which preserves the homogeneity and the isotropy. 
We choose the Robertson-Walker metric 
\begin{eqnarray}
ds^2 &=& -N(t)^2dt^2+a(t)^2h_{ij}(x)dx^idx^j. 
   \label{ansatz HL}
\end{eqnarray}
Here we consider the so-called projectable theory. 
The spatial metric $g_{ij}=a^2h_{ij}$ satisfies 
\begin{equation}
R_{ij} = 2ka^{-2}g_{ij}. 
\end{equation}
Therefore, the covariant derivative $\nabla_i$ of the Ricci tensor vanishes. 
In addition, since $h_{ij}$ is conformally flat, the Cotton tensor $C^{ij}$ for $h_{ij}$ vanishes. 
Inserting the metric ansatz into the action, we obtain the following reduced action: 
\begin{equation}
S_R = V\int dt\sqrt{g}N\left[ \frac{\alpha}{N^2}(3-9\lambda)\left(\frac{\dot{a}}a \right)^2+\frac{12k^2(\zeta+3\eta)}{a^4}
 +\frac{6k\xi}{a^2}+\sigma+L_m \right].
\end{equation}
% Note: 
% \begin{eqnarray}
% K_{ij} &=& \frac1N\frac{\dot{a}}{a}g_{ij}, \\
% K_{ij}K^{ij} &=& \frac3{N^2}\left( \frac{\dot{a}}a \right)^2, \\
% K^2 &=& \frac9{N^2}\left( \frac{\dot{a}}a \right)^2, \\
% R_{ij}R^{ij} &=& \frac{12k^2}{a^4}, \\
% R &=& \frac{6k}{a^2}, \\
% R^2 &=& \frac{36k^2}{a^4}. 
% \end{eqnarray}

We assume that the matter consists of a dust. 
The coupling of the dust to the metric is determined by 
\begin{eqnarray}
\delta_g\int d^4x\sqrt{-g}L_m 
&=& \frac12\int d^4x\sqrt{-g}\delta g_{\mu\nu}T^{\mu\nu} \nonumber \\
&=& \frac12\int d^3xdt\sqrt{g}N\Bigl[ \delta N(-2NT^{00})+\delta N^i(2g_{ij}N^jT^{00}+2g_{ij}T^{j0}) \nonumber \\
& &+\delta g_{ij}(N^iN^jT^{00}+N^jT^{i0}+N^iT^{j0}+T^{ij}) \Bigr], 
\end{eqnarray}
where 
\begin{equation}
T^{00} = \rho(t), \hspace{5mm} T^{i0} = 0 = T^{ij}. 
\end{equation}

The equation of motion of $N$ is 
\begin{equation}
3\alpha(1-3\lambda)\left( \frac{\dot{a}}a \right)^2 = \frac{12k^2(\zeta+3\eta)}{a^4}+\frac{6k\xi}{a^2}+\sigma-\rho. 
   \label{Friedmann}
\end{equation}
Here we chose $N=1$. 

A remark on the derivation of (\ref{Friedmann}) is in order. 
Since we assumed that $N$ depends only on $t$, the equation of motion of $N$ does not provide a local equation in general. 
Instead, one obtains 
\begin{eqnarray}
0 
&=& \int d^3x\sqrt{g}\Bigl[ -\alpha(K_{ij}K^{ij}-\lambda K^2)+\beta C_{ij}C^{ij}+\gamma{\cal E}^{ijk}R_{il}\nabla_jR^l{}_l 
    \nonumber \\
& & \hspace*{1.5cm} +\zeta R_{ij}R^{ij}+\eta R^2+\xi R+\sigma +L_m \Bigr]. 
\end{eqnarray}
In the present case, however, the spatial dependence of the integrand disappears due to the ansatz (\ref{ansatz HL}), 
and therefore, the integral provides just a trivial 
volume factor. 
In this way, one can obtain an equation which looks local, enabling one 
to obtain the above Friedmann-like equation (\ref{Friedmann}). 

The matter density $\rho$ is given in terms of $a$ through the conservation law. 
The invariance under the time reparametrization provides 
\begin{equation}
\partial_t[a^3\rho] = 0. 
   \label{conserve}
\end{equation}
Note that, as in the case for 
the equation of motion of $N$, the conservation law for the time reparametrization is not a local equation 
in general, since Horava-Lifshitz gravity preserves only a restricted part of the diffeomorphisms. 
In the case here, the ansatz of the homogeneity and the isotropy enables one to obtain (\ref{conserve}). 
The equation (\ref{conserve}) implies 
\begin{equation}
\rho(t) = \frac{\rho_0}{a(t)^{3}}, 
\end{equation}
As in the relativistic case, we rescale the radial coordinate $r$ so that $a(0)=1$. 

Now the equation (\ref{Friedmann}) becomes 
\begin{eqnarray}
\left( \frac{da}{dt} \right)^2 &=& -V(a),  
    \label{F2} \\
V(a) &=& -\frac{c_{-2}}{a^2}-\frac{c_{-1}}{a}-c_0-c_2a^2,
\end{eqnarray}
where 
\begin{eqnarray}
c_{-2} &=& \frac{4k^2(\zeta+3\eta)}{\alpha(1-3\lambda)}, \\
c_{-1} &=& -\frac{\rho_0}{3\alpha(1-3\lambda)}, \\
c_0 &=& \frac{2k\xi}{\alpha(1-3\lambda)}, \\
c_2 &=& \frac{\sigma}{3\alpha(1-3\lambda)}. 
\end{eqnarray}
We consider a generic case where $\zeta+3\eta\ne0$. 
This is the case when the theory satisfies the detailed balance condition, and a small deviation from the detailed balance may 
not change the sign of $\zeta+3\eta$. 

We choose the initial condition $\dot{a}(0)=0$, $a(0)=1$ as in the relativistic case. 
This implies 
\begin{equation}
c_{-2}+c_{-1}+c_0+c_2 = 0, 
   \label{initial}
\end{equation}
which determines $k$. 
This equation is second order in $k$, while $k$ is assumed to be real. 
A real $k$ is allowed iff 
\begin{equation}
4(\zeta+3\eta)(\rho_0-\sigma)+3\xi^2 \ge 0. 
    \label{existence}
\end{equation}

In general, there is a singular situation\footnote{
We would like to thank Soo-Jong Rey for pointing out this singular situation.
}. 
If $\rho_0=\sigma$ is satisfied, then $k=0$ is a solution of (\ref{initial}). 
In this case, $c_{-2}$ and $c_0$ in $V(a)$ vanish. 
As long as $\lambda>\frac13$, the resulting equation is the same as (\ref{relF}) up to a rescaling of $t$, implying that 
the dust collapses completely and a singularity is formed. 
Note that the formation of the singularity in this case should be exceptional due to the highly symmetric setup we assumed. 
The solution with $k=0$ has the spatial hypersurface which is flat, and the higher derivative terms do not contribute to the 
dynamics of $a$. 
This is due to the cancellation between the positive energy $\rho_0$ and the negative energy $-\sigma$. 
In general, a small deviation from the flat spatial hypersurface 
would induce a small but non-zero $c_{-2}$, at least effectively, and 
therefore, the behavior of the solution would be similar to the one with $k\ne0$ which will be investigated below. 

\vspace{5mm}

\begin{figure}
\epsfxsize=2.2in
\begin{center}
\includegraphics[scale=0.5]{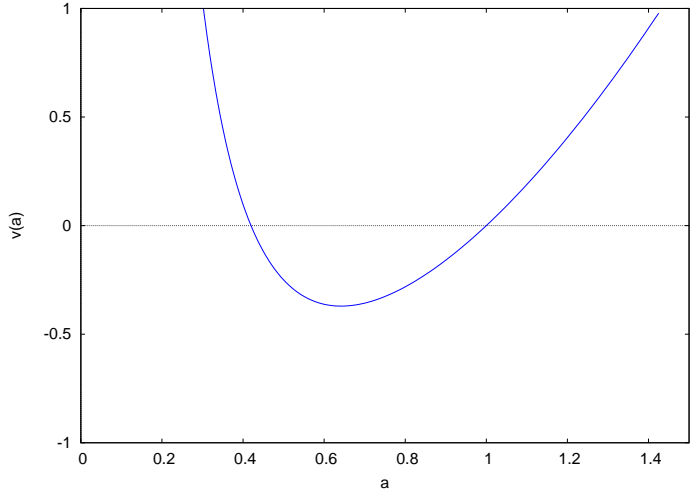}
\end{center}
\caption{The plot of $v(a)=\frac{16(3\lambda-1)^2}{\kappa^4\mu^2\Lambda^2}V(a)$ with $\tilde{k}=\frac12$. }
\label{V(a)}
\end{figure}

Let us consider the case in which the detailed balance condition is satisfied. 
In this case, the coefficients in the action (\ref{HLgravity}) are parametrized as follows \cite{Kiritsis:2009sh}:  
\begin{eqnarray}
\alpha &=& \frac2{\kappa^2}, \\
\zeta &=& -\frac{\kappa^2\mu^2}8, \\
\eta &=& \frac{\kappa^2\mu^2}{8(1-3\lambda)}\frac{1-4\lambda}4, \\
\xi &=& \frac{\kappa^2\mu^2}{8(1-3\lambda)}\Lambda, \\
\sigma &=& \frac{\kappa^2\mu^2}{8(1-3\lambda)}(-3\Lambda^2).
\end{eqnarray}
They give, for example, $c_{-2}=-\frac{k^2\kappa^2\mu^2}{16(3\lambda-1)^2}$. 
Using those parametrization and after a suitable rescaling of $t$, the equation (\ref{F2}) becomes  
\begin{equation}
\left( \frac{da}{dt} \right)^2 = -\left( \frac{\tilde{k}}{a}-a \right)^2
 +\frac83\frac{(3\lambda-1)\rho_0}{\kappa^2\mu^2\Lambda^2}\frac1a, 
   \label{F3}
\end{equation}
where $\tilde{k}=\frac k{|\Lambda|}$. 
%\begin{eqnarray}
%c_{-2} &=& -\frac{\kappa^4\mu^2}{16(1-3\lambda)^2}k^2, \\
%c_{-1} &=& -\frac{\kappa^2\rho_0}{6(1-3\lambda)}, \\
%c_0 &=& \frac{\kappa^4\mu^2}{8(1-3\lambda)^2}k\Lambda, \\
%c_2 &=& -\frac{\kappa^4\mu^2}{16(1-3\lambda)^2}\Lambda^2. 
%\end{eqnarray}
The condition (\ref{existence}) for the existence of a solution reduces to $\lambda>\frac13$, and the solutions 
of (\ref{existence}) are 
\begin{equation}
\tilde{k} = 1\pm\sqrt{\frac{8(3\lambda-1)}{3\kappa^2\mu^2\Lambda^2}\rho_0}. 
\end{equation}
In the following, we concentrate on the case $\lambda>\frac13$. 
Typical shape of $V(a)$ is depicted in Fig. \ref{V(a)}. 

It turns out that $V(a)$ is a concave function of $a$, and the equation $V(a)=0$ has two real solutions. 
We expect that the dust would contract at first, and this is compatible with (\ref{F3}) only if $V'(1)>0$. 
The value of $V'(1)$ is 
\begin{equation}
V'(1) = \frac{\kappa^4\mu^2\Lambda^2}{16(1-3\lambda)^2}(\tilde{k}+3)(1-\tilde{k}), 
\end{equation}
and this quantity can be positive iff $-3<\tilde{k}<1$, implying  
\begin{equation}
\rho_0 < \rho_{\rm max} = \frac{6\kappa^2\mu^2\Lambda^2}{3\lambda-1}. 
\end{equation}
This condition shows that there exists a maximum possible energy density determined by the parameters of the theory. 
If the initial density is above the maximum, the dust would start {\it expanding} rather than contracting. 

\begin{figure}
\epsfxsize=2.2in
\begin{center}
\includegraphics[scale=0.5]{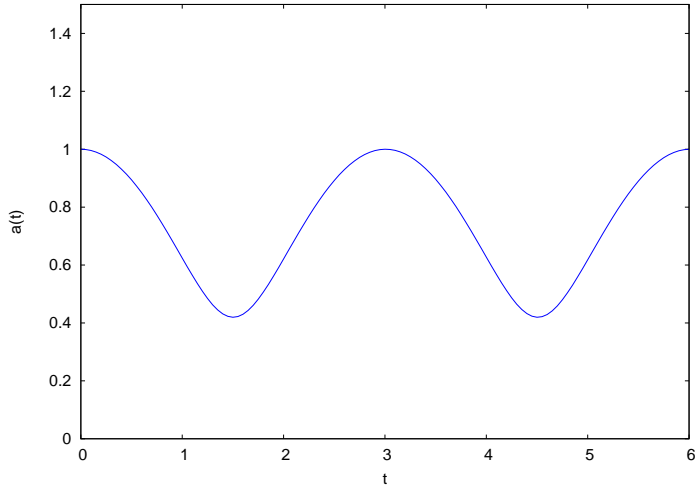}
\end{center}
\caption{The time evolution of the scale factor $a(t)$ for $\tilde{k}=\frac12$. }
\label{a(t)}
\end{figure}

The qualitative behavior of the solution of (\ref{F2}) can be easily understood through the analogy with the corresponding classical 
mechanical system in which $a$ is a coordinate variable: 
\begin{equation}
\left( \frac{da}{dt} \right)^2+V(a) = 0. 
   \label{mechanical}
\end{equation}
Suppose that $\rho_0<\rho_{\rm max}$ is satisfied. 
The value of $a$ starts decreasing at first, due to the gravitational interaction, and the energy density becomes large. 
However, the term $\frac{\tilde{k}^2}{a^2}$ in (\ref{F3}), which is absent in the relativistic case (\ref{relF}), 
acts as a {\it repulsive} force, 
and it prevents the dust from collapsing completely. 
Instead, the system will attain a minimum value of $a$ at some finite time, and then expand again. 
The expansion will also stop in a finite time, and the system will oscillate with a finite period. 
A numerical solution for $\tilde{k}=\frac12$ is shown in Fig.\ref{a(t)}. 

Note that the appearance of this periodic behavior would be due to the fact that we have considered a highly symmetric 
situation. 
If there would be a small asymmetry, then the oscillation of the dust would probably 
emit the gravitational wave, and the system would 
settle down to a static configuration; a compact ball of the dust with a finite size. 
The energy density would be also finite during the process. 

\vspace{5mm}

\begin{figure}
\epsfxsize=2.2in
\begin{center}
\includegraphics[scale=0.5]{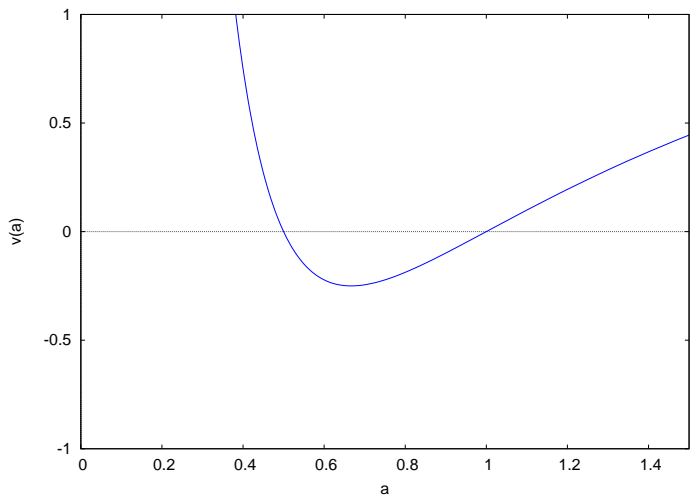}
\end{center}
\caption{The plot of $v(a)$ with $\tilde{\sigma}=1$ and $\tilde{k}=\frac12$. }
\label{beta=1}
\end{figure}

It is fairly obvious that the qualitative behavior of the dust discussed above is preserved by a small perturbation 
of the theory away from the detailed balance, as long as $c_{-2}$ is kept negative. 
For example, the addition of a term proportional to $R$ simply shifts the ``energy'' of the mechanical system (\ref{mechanical}), 
which will not change the behavior of the solution as long as the condition $V(a)=0$ has two real solutions for $a$. 
Another simple deviation from the detail balance case can be achieved by parametrizing $\sigma$ as 
\begin{equation}
\sigma = \frac{\kappa^2\mu^2}{8(1-3\lambda)}(-3\lambda^2)(1-{\tilde{\sigma}}), 
\end{equation}
with another parameter $\tilde{\sigma}$. 
The equation (\ref{F3}) becomes 
\begin{equation}
\left( \frac{da}{dt} \right)^2 = -\left( \frac{\tilde{k}}{a}-a \right)^2
 +\frac83\frac{(3\lambda-1)\rho_0}{\kappa^2\mu^2\Lambda^2}\frac1a+{\tilde{\sigma}} a^2. 
    \label{F4}
\end{equation}
It turns out that the behavior of the solution is qualitatively the same as the one for $\tilde{\sigma}=0$, as long as $\tilde{\sigma}<1$ is 
satisfied. 
Let us consider the case $\tilde{\sigma}=1$ in which the cosmological constant $\sigma$ vanishes. 
The initial condition for $a$ determines $\tilde{k}$ as 
\begin{equation}
\tilde{k} = 1- \sqrt{1+\frac83\frac{(3\lambda-1)\rho_0}{\kappa^2\mu^2\Lambda^2}}, 
\end{equation}
which is compatible with the condition $V'(1)>1$, or equivalently, $-2<\tilde{k}<0$. 
The shape of $V(a)$ with $\tilde{k}=-1$ is depicted in Fig. \ref{beta=1}. 
It is concluded that even in the case $\tilde{\sigma}=1$, there is a solution for the dust which is oscillating. 
Notice that there is a crucial difference from the result in general relativity, in addition to the absence of the singularity 
formation. 
The value of $k$ is always positive in general relativity without cosmological constant, but in Horava-Lifshitz gravity $k$ is 
always {\it negative} if the cosmological constant vanishes. 
This difference may affect the space-time geometry outside the dust through the continuity of the metric at the surface of the 
dust. 
Therefore, the difference between general relativity and Horava-Lifshitz gravity might be observed by an observer at infinity. 
If $\tilde{\sigma}$ is sufficiently large, there is no solution in which the dust starts contracting at first. 
This is because of the expansion due to the large positive cosmological constant. 

\vspace{5mm}

We have considered the dynamics of the dust satisfying the initial condition $a(0)=1,\dot{a}(0)=0$. 
It is very easy to generalize the above analysis to the case in which $\dot{a}(0)=0$ is still assumed but 
$a(0)$ can be arbitrary. 
This generalization is necessary if one would like to consider a realistic situation where the higher-derivative terms are 
small at the beginning of the collapse. 

Let us define the normalized scale factor $\bar{a}$, 
\begin{equation}
\bar{a}(t) := \frac{a(t)}{a(0)}. 
\end{equation}
In terms of $\bar{a}$, the equation (\ref{F4}) becomes 
\begin{equation}
\left( \frac{d\bar{a}}{dt} \right)^2 = -\left( \frac{\tilde{k}/a(0)^2}{\bar{a}}-\bar{a} \right)^2
 +\frac83\frac{(3\lambda-1)}{\kappa^2\mu^2\Lambda^2}\frac{\rho_0/a(0)^3}{\bar{a}}+\tilde{\sigma} \bar{a}^2. 
\end{equation}
The effect of the generic initial value of $a(t)$ appears as a simple rescaling of $\tilde{k}$ and $\rho_0$. 
It is easy to see that the qualitative behavior of the dust does not change by varying the value of $a(0)$. 
If $a(0)$ would be taken to be large, then $\tilde{k}/a(0)^2$ would be negligible at first, and the behavior of the dust is 
well approximated by the collapse in Einstein gravity with a negative cosmological constant reviewed at the beginning of 
this section. 
When $\bar{a}$ becomes small enough so that the effect of $\tilde{k}$ becomes relevant, the dynamics of the dust deviates from 
the complete collapse in Einstein gravity, and turns into expansion.

\vspace{1cm}

\section{Discussion}

\vspace{5mm}

We have discussed the dynamics of a matter in Horava-Lifshitz gravity. 
It would be natural to expect that the modification of the UV behavior of the gravity part should be accompanied by a similar 
modification for the matter part. 
We investigated the propagation of a scalar field governed by such a modified action in the WKB approximation. 
The consistency of the scalar propagation with the causality restricts the form of the action. 
For the consistent scalar field, there is no relevance of the light cone. 
Therefore, in contrast to general relativity, a formation of a singularity would directly imply the breakdown of the 
predictability. 
We also investigated the dust collapse in Horava-Lifshitz gravity. 
We found that any singularity would not be formed during the dust collapse preserving the homogeneity and the isotropy 
which is quite different from the situation in general relativity. 
It is tempting to speculate that the formation of any singularity is forbidden in Horava-Lifshitz theory, 
but the investigation on this issue is left for further research. 
The breakdown of the predictability due to a singularity formation would not be realized in 
Horava-Lifshitz gravity if the formation of the singularity would be forbidden. 

The repulsive interaction which prevents the dust collapse is induced by the higher derivative terms. 
However, it is clear that every higher derivative terms 
do not necessarily 
provide such a repulsive interaction. 
In fact, one can choose the action so that $c_{-2}$ in (\ref{F2}) is positive. 
The negative sign of $c_{-2}$ should be related to the fact that the ``potential term'' of Horava-Lifshitz gravity is positive 
definite in the detailed balance case. 
Due to this structure, a high-curvature configuration costs much energy than a low-curvature one, and therefore the former is not 
favored. 
It seems that the choice of $c_{-2}$ might be restricted since $c_{-2}$ could be related to the high energy behavior of the 
graviton propagator. 
It is interesting to investigate this point further. 

One interesting issue to be studied next would be the outer solution for the dust collapse. 
In this paper, we only considered the solution which describes the inside of the dust. 
If the dust is a ball with a finite radius, then the metric describing the outside of the ball would be spherically symmetric but 
is not homogeneous. 
In general relativity, one can find the solution describing the outside of the dust, thanks to Birkhoff theorem. 
It is very interesting to study whether there is such a theorem in Horava-Lifshitz gravity. 
Since the available diffeomorphisms in Horava-Lifshitz gravity are less than those in general relativity, such a theorem, if 
exists, must be a much weaker one. 

Another interesting issue would be the gravitational radiation in Horava-Lifshitz gravity. 
As we mentioned at the end of the previous section, the construction of a more realistic model of the dust collapse might require 
the understanding of the emission of the gravitational wave from the oscillating matter. 
From the physical ground, the spherically symmetric wave should be forbidden also in Horava-Lifshitz gravity as long as it is 
assumed to approach general relativity in low energy. 
Since otherwise the emitted gravitational 
wave at a high energy region near the center may propagate toward the outer low energy region where 
such a wave cannot exist as in general relativity. 
If a spherically symmetric wave does not exist, then the variety of the spherically symmetric solutions would be rather limited, 
and a weaker version of Birkhoff theorem could be expected. 
It is interesting to clarify this issue.

\vspace{2cm}

{\bf \Large Acknowledgements}

\vspace{5mm}

We would like to thank Soo-Jong Rey and Satoshi Yamaguchi for valuable discussions and comments. 
This work was supported by the BK21 program of the Ministry of Education, Science and Technology, 
National Science Foundation of Korea Grants R01-2008-000-10656-0, 
2005-084-C00003, 2009-008-0372 and EU-FP Marie Curie Research 
\& Training Network HPRN-CT-2006-035863 (2009-06318).

\end{document}